\documentclass[pra,showpacs,floatfix]{revtex4-1}% 

\usepackage{amssymb}
\usepackage{amsmath}
\usepackage{bm}
\usepackage{afterpage}
\usepackage{graphicx}

\def\beq{\begin{equation}}
\def\eeq{\end{equation}}

\def\om{\omega}
\def\a{\alpha}
\def\b{\beta}

\def\s{\sigma}
\def\D{\Delta}

\def\ad{a^\dagger}

\def\p{\phi}

\def\ra{\rightarrow}

\def\Cc{\mathbb{C}}
\def\Zz{\mathbb{Z}}
\def\Rr{\mathbb{R}}
\def\Zpl{\rm{I\!N}}

\def\Hpm{{\cal H}_\pm}
\def\sH{\tilde{H}}
\def\ua{\!\!\uparrow}
\def\da{\!\!\downarrow}

\def\B{{\cal B}}

\def\Prj{\hat{P}}
\def\ph{\phantom}

\begin{document}

  \title{Continued Fractions and the Rabi Model}

  \author{Daniel Braak}
  \email{daniel.braak@physik.uni-augsburg.de} 
  \address{EP VI and
   Center for Electronic Correlations and Magnetism,\\
Institute of Physics,
    University of Augsburg, 86135 Augsburg, Germany}
  %\date{November 12, 2012}

  \begin{abstract}
  Techniques based on continued fractions to compute numerically the 
spectrum of the quantum Rabi model are reviewed. They are of two 
essentially different types. In the first case,
the spectral condition is implemented using a representation
in the infinite-dimensional Bargmann space of analytic functions. This approach is shown to
approximate the correct spectrum of the full model if the continued fraction is
truncated at sufficiently high order. 
In the second case, one considers the limit of a sequence of models defined in
finite-dimensional state spaces. Contrary to the first, the second approach is ambiguous and
can be justified only through recourse to the analyticity argument from the first method.

  \end{abstract}


  \maketitle

\section{Introduction}

The fully quantized Rabi model \cite{rabi,JC} constitutes probably
the simplest strongly coupled quantum system with an infinite
dimensional Hilbert space. It is described by the Hamiltonian ($\hbar=1$),
\beq
H_R= \om\ad a +g\s_x(a+\ad) +\D\s_z. 
\label{ham1}
\eeq
$H_R$ possesses a $\Zz_2$-symmetry which renders it integrable \cite{db}, 
although the associated two invariant subspaces (parity-chains) are
each infinite-dimensional themselves. 
The regular spectrum of (\ref{ham1}) is determined in each parity chain 
by the zeroes of a
transcendental function $G_\pm(x)$, which can be expressed through
confluent Heun functions \cite{ron}.  
$G_\pm(x)$  has poles at integer multiples of $\om$, which allows to obtain results on the distribution of eigenvalues and the location of
degeneracies which are related to the quasi-exact (exceptional) spectrum \cite{db}. 
The approach based on the symmetry of the model leads in this way to a 
unified picture of all spectral properties.

Independent from the qualitative understanding of the physics governing the
quantum Rabi model is the question whether the analytical solution
could invalidate the widely used {\it numerical} determination of the
spectrum through exact diagonalization on a truncated state space, or
the equivalent continued fraction techniques 
\cite{schweb,swain,sten,reik,klen,reik2,szopa,mol}.
There has been recent confusion on this point  
\cite{mor,zie} and the present paper intends to 
clarify the situation
regarding the methods based on continued fractions.
These fall into two classes, which we denote with A and B. 

\section{Method A}

One considers the full model, i.e. the state space is assumed
to be infinite-dimensional, ${\cal H}=L^2(\Rr)\otimes\Cc^2$. 
Work in this direction goes back to the 
seminal paper by Schweber \cite{schweb}.
He used the isomorphism between $L^2(\Rr)$ and the Bargmann space $\B$ of analytic functions \cite{barg} to obtain a differential equation for the spin-up component of the eigenfunction  
with energy $E$, 
\beq
\psi_E(z)=e^{-gz/\om}\sum_{n=-\infty}^\infty K_n(E)\left(z+\frac{g}{\om}\right)^n,
\label{expan}
\eeq 
leading to a
linear three-term recurrence relation for the coefficients $K_n(E)$,
\beq
nK_{n}=f_{n-1}(E)K_{n-1}-K_{n-2},
\label{recur}
\eeq
 with
 \beq
f_n(E)=\frac{2g}{\om}+\frac{1}{2g}
\left(n\om-x(E) +\frac{\D^2}{x(E)-n\om}\right)
\label{f-n}
\eeq
and $x(E)=E+g^2/\om$.
The same equation results if the $\Zz_2$-invariance is used to project the Hamiltonian to one of its invariant subspaces with fixed parity \cite{db}.
Each of the parity-chains $\Hpm$ is isomorphic to $\cal B$ \cite{db} and we set ${\cal H} ={\cal H}_+={\cal B}$ in the following, confining the discussion to positive parity. The energy eigenvalues $E_n$, $n=0,1,2,\ldots$ of $H_R$ are determined by imposing on $\psi_E(z)$ the condition to be an element of $\cal B$, henceforth called the ``spectral condition''. This condition consists of two parts, the first is the familiar normalizability, $|\psi_E(z)|<\infty$, where $|\cdot|$ denotes the norm in $\cal B$. Because the differential equation for $\psi_E(z)$ in (\ref{expan}) has an irregular singular point of rank one at infinity \cite{db2},  $\psi_E(z)\sim e^{cz}z^\rho(c_0+c_1/z+\ldots)$ with  complex constants $c,\rho,c_0,c_1$ \cite{ince}. As $\cal B$ contains all functions growing like $e^{cz}$ for $|z|\ra\infty$, $\psi_E(z)$ fulfills the first part of the spectral condition for any real $E$. The second part demands $\psi_E(z)$ to be {\it analytic} everywhere in $\Cc$, and it is this second part which fixes the discrete set $\{E_n\}_{n\in\Zpl_0}$, the spectrum of $H_R$. The differential equation for $\psi_E(z)$ has two regular singular points 
at $z=\pm g/\om$ and $\psi_E(z)$ will be analytic in $\Cc$ if it is analytic at both of these points. $\psi_E(z)$ is expanded  in (\ref{expan})  around $z=-g/\om$ and  analytic there if $K_n=0$ for $n\le -1$. This leads to
the initial condition for the recurrence (\ref{recur}): $K_0=1$ and
$K_1(E)=f_0(E)$. It yields still no equation for $E$, which is obtained by demanding analyticity at the second singular point $z=g/\om$.  
From the
Poincar\'e analysis of (\ref{recur}) it follows that there are two possibilities for the limiting behavior of the $K_n(E)$:
$K_{n+1}(E)/K_n(E)\ra \om/(2g)$ or $K_{n+1}(E)/K_n(E)\ra 0$. The latter behavior
corresponds to the unique {\it minimal} solution $\{K^{min}_n\}_{n\in\Zpl}$ of (\ref{recur})  with initial condition $K_0=1$ \cite{note1}. 
Note that $K^{min}_1$ is determined
in terms of $\{f_1(E),f_2(E),\ldots\}$ for any $E$ and does not depend on $f_0(E)$ \cite{gaut,erde}. 
All other solutions $\{K_n\}_{n\in\Zpl}$ of (\ref{recur}) are {\it dominant}, i.e. $\lim_{n\ra\infty} K^{min}_n/K_n=0$.
In general, $K_1^{min}(E)\neq f_0(E)$, but for $E\in\{E_0,E_1,\ldots\}$ one has
$K_1^{min}(E_n)=f_0(E_n)$. This condition  determines the spectrum because only 
then $\psi_{E_n}(z)$ is analytic at both $-g/\om$ and $g/\om$ and therefore in all of $\Cc$. Analyticity at $-g/\om$ fixes $K_1=f_0(E)$ and analyticity at $g/\om$
enforces an infinite radius of convergence
of the expansion (\ref{expan}), which
would be finite ($R=2g/\om$) for any dominant solution with $K_{n+1}(E)/K_n(E)\ra \om/(2g)$. Only the minimal solution
yields $R=\infty$, therefore $K_1(E)=K_1^{min}(E)=f_0(E)$, which is the sought after equation for $E$. 
 
To compute the minimal solution of (\ref{recur}),
Schweber proceeds to represent the quotient $K_1(E)/K_0$ through the
infinite  continued
fraction  
\beq
\frac{K_1(E)}{K_0}=f_0(E)=1|f_1(E)-2|f_2(E)-3|f_3(E)-\ldots.
\label{cont-frac2}
\eeq
As pointed out in \cite{db}, this equation amounts to a tautology if the formal
expression on its right hand side is not augmented with a  prescription for the asymptotics
of the tail $\xi_N(E)$ in
\beq
F_\infty(E)=1|f_1(E)-2|f_2(E)-3|f_3(E)-\ldots 
=1|f_1(E) - 2|f_2(E)-\ldots (N-1)|f_{N-1}(E)-\xi_N(E). 
\eeq
Fortunately, the standard definition of an infinite continued fraction 
$F_\infty(E)$ as the
limit (if it exists) of a sequence of finite continued fractions $F_n(E)$,
\beq
F_\infty(E)=\lim_{n\ra\infty}
\Big(1|f_1(E)-2|f_2(E)-\ldots n|f_n(E)\Big) 
= \lim_{n\ra\infty}F_n(E)  
\label{cf-limit}
\eeq
yields the correct prescription. The reason is that
the limit (\ref{cf-limit}) exists in case of the asymptotic behavior
\beq
\lim_{n\ra\infty} \frac{\xi_n(E)}{f_{n-1}(E)}=0
\eeq
but because $\xi_n=nK_n/K_{n-1}$ and $f_n\sim n\om/(2g)$ for large $n$, this can only happen if
$\lim_{n\ra\infty}K_n(E)/K_{n-1}(E)=0$, which characterizes the minimal solution. Pincherle's theorem, which connects minimal solutions with $K_0\neq 0$ to continued fractions is a simple consequence of the uniqueness of $\{K_n^{min}\}$ \cite{gaut}. 
In other words, the continued fraction (\ref{cf-limit}) is a short-hand notation for
the minimal solution $K_1^{min}(E)/K_0$ of the recurrence (\ref{recur}), which is
defined for arbitrary $E$. 
Schweber has proven the (pointwise) convergence  of (\ref{cf-limit}) 
 using a 
corollary of Pringsheim's main theorem \cite{wall,per}. However, in contrast to
power series expansions, the actual value of (\ref{cf-limit}) can only be 
obtained by checking convergence numerically for each $E$ separately, by computing
a set of $F_n(E)$ with large enough $n$. It is not possible to deduce from the
smallness of the tail
\beq
\xi_n(E)=n|f_n(E)-(n+1)|f_{n+1}(E)- \ldots  
\eeq
the value of $F_\infty(E)$ for given $F_{n-1}(E)$  to a prescribed accuracy 
because the continued fraction itself may diverge while $\xi_n$ stays 
finite \cite{note2}.
Therefore, the actual computation concerns always some $F_N(E)$ with finite $N$.
It is easy to see that
the equation $f_0(E)=F_N(E)$ is just the spectral condition
for the Hamiltonian $H_N=\hat{P}_NH_R\hat{P}_N$, where $\hat{P}_N$ denotes
the projector onto
the finite-dimensional Hilbert space ${\cal H}_N$, spanned by 
functions of the form
\beq
 \p_n(z) = e^{-gz/\om}\left(z+\frac{g}{\om}\right)^n, \qquad n=0,1,2,\ldots N.
\label{expanf}
\eeq 
If $\psi_E(z)=\sum_{n=0}^\infty K_n\p_n(z)$ satisfies formally $(H_R-E)\psi_E(z)=0$, it follows
\beq
\sum_{n=0}^\infty K_n H_R\p_n= 
\sum_{n=0}^\infty \big[(2gf_n(E)+E)K_n-2g[(n+1)K_{n+1}+K_{n-1}]\big]\p_n,
\label{expan3}
\eeq
because of (\ref{recur}). Obviously, $\Prj_NH_R\Prj_N\p_{N+1}=0$, which entails that the coefficient of
$\p_N$ in (\ref{expan3}) becomes $(2gf_N(E)+E)K_N-2gK_{N-1}$, which means $K_{N+1}=0$, i.e. the recurrence
(\ref{recur}) is cut off at $K_N$. Now (\ref{recur}) together with the initial condition $K_1=f_0(E)$  and additionally $K_{N+1}=0$ is equivalent to an equation involving a finite continued fraction,
\beq
f_0(E)=1|f_1(E)-2|f_2(E)-3|f_3(E)-\ldots N|f_N(E) = F_N(E),
\label{spec-fin}
\eeq
and provides therefore the spectral condition in the space ${\cal H}_N$.

Note that this space does not correspond to the span of a finite subset of 
Fock states $\{z^{j_n},n=0\ldots N\}$ as the truncated spaces ${\cal H}^{(N)}$ 
used in method B, but all elements of ${\cal H}_N$ can be approximated to arbitrary precision by elements of
 ${\cal H}^{(N)}$ for sufficiently large $N$
(see below). 
Each solution $E_n^N$ of (\ref{spec-fin}) approximates some solution $E_n$ of $F_\infty(E)=f_0(E)$ to  precision 
$\varepsilon$ if
$N > {\cal N}(n,\varepsilon)$. 
The latter property follows from the fact that the convergence of the tail $\xi_N(E)$ depends on the value of
$E$ being smaller than a certain upper bound depending on $N$ (see \cite{schweb} or Eq.~(\ref{bound}) below).
The solution $E^N_n$  of (\ref{spec-fin}) must be smaller than this bound for $F_N(E^N_n)$ to be a good approximation of $F_\infty(E^N_n)$. However, for each finite $E_n$ being a solution of the exact equation $f_0(E)=F_\infty(E)$, there is a $N(E_n)$ such that for all $N>N(E_n)$ the function $F_N(E_n)$ approximates $F_\infty(E_n)$ to a given precision
and $E_n\approx E^N_n$ will be thus close to the solution $E^N_n$ of  (\ref{spec-fin}).
 
One may formalize this statement as follows. The domain of the operators $H_N$ is extended to $\cal H$ by setting $H_N\p_m=0$ for $m\ge N+1$. Then the spectrum $\s(H_R)=\{E_n\}_{n\in\Zpl_0}$ of the operator $H_R$ is the limit of the spectra $\s(H_N)=\{E_n^N\}_{n\in\Zpl_0}$ in the sense that there exist a number $M(N)$, unbounded in $N$, such that
\beq
\lim_{N\ra\infty} \max_{n<M(N)}|E_n^N-E_n| = 0.
\label{cond-lim}
\eeq    
The spectral condition on ${\cal H}_N$
provides an approximation to the spectrum of the full model precisely
because it approximates the {\it exact} condition of analyticity in the Bargmann space.
It follows that the continued fraction approach initiated by Schweber and pursued by
several authors \cite{sten,reik,klen,reik2,szopa} 
can be used to obtain
the spectrum of
the quantum Rabi model to arbitrary precision. The method does not use  the
$\Zz_2$-symmetry of $H_R$ as the Bargmann condition is implemented via
(\ref{cont-frac2}).
Indeed, (\ref{cont-frac2}) only depends on $\D^2$ and does not discern
the two parity chains, which is the reason why
the relation of isolated (Juddian) solutions at $x(E)=n\om$ 
to level crossings, i.e. reducible representations of $\Zz_2$, appears
to be accidental within Schwebers approach \cite{kus-lew}. 
The infinite continued fraction $F_\infty(E)$  
is completely opaque to an analysis of its general behavior as function of $E$: 
$F_\infty(E)$ is the formal quotient of two functions,
\beq
F_\infty(E)=\lim_{n\ra\infty}\frac{A_n(E)}{B_n(E)}
\eeq
where $A_n,B_n$ both satisfy the recurrence relation
\beq
C_n=f_n(E)C_{n-1}-nC_{n-2},\quad n\ge 2
\eeq
and $A_0=0,A_{-1}=1,B_0=1,B_{-1}=0$.
Only the quotient of $A_n$ and $B_n$ is well-defined for $n\ra\infty$. 
Therefore the location of zeros and poles of $F_\infty(E)-f_0(E)$
cannot be read off from the behavior of numerator or denominator in the
limit $n\ra\infty$, in contrast to the function $G_\pm(x)$ 
whose pole structure is known and leads to simple rules for the
distribution of eigenvalues. Method A can be used to calculate the
spectrum of the Rabi model to any desired accuracy, but provides
no  insight beyond  direct
numerically exact diagonalization in a finite-dimensional state space
with dimension $N+1$, if (\ref{cont-frac2}) is truncated at the
order $N$.  

\section{Method B}

This approach goes back to Swain \cite{swain}. It is based on a
three-term recurrence relation for the determinants of tridiagonal matrices.
Because the quotient of two such determinants can be interpreted as a matrix
element of the resolvent of $H_R^{(N)}$, if defined on a truncated Hilbert space, ${\cal H}_t^{(N)}=\Cc^{N+1}\otimes\Cc^2$, one
may write this matrix element as a finite continued fraction. 
By repeating Swain's calculation, the authors of recent work \cite{mol,zie}
 tie the ``solvability" of the Rabi model to his representation of the resolvent.
 It is further claimed in \cite{zie} that this representation, which is defined in ${\cal H}_t^{(N)}$
 only, has a well-defined limit $N\ra\infty$ at least for $g^2<\om$ and provides in this
 way an exact solution of the Rabi model {\it without} making use of the analyticity condition
 employed by method A. This claim is based on the belief that the operator $H_R$,
defined on an infinite-dimensional Hilbert space, can be approximated by the sequence of
finite-dimensional matrices $H_R^{(N)}$, which is incorrect because $H_R$ is not compact \cite{reed}.
Nevertheless, the full spectrum of $H_R$ can be obtained by method B as well ---
but this is entirely due to the equivalence  of the spaces ${\cal H}_t^{(N)}$ to the
spaces ${\cal H}_N\otimes\Cc^2$ from method A --- and rests therefore again on the
Bargmann criterion, as shown in the following.     

There are several equivalent formulations of Swain's approach \cite{swain,mol,zie,durst}.   
In its simplest form \cite{mol}, one considers the Hamiltonian in each parity chain
 separately, where it corresponds to a tridiagonal matrix. Define the parity chain $\Hpm=$ span$\{|n,\pm(-1)^n\rangle\}_{n\in\Zpl_0}$
 with $|n,1\rangle=|n\rangle\otimes|\ua\rangle$  and $|n,-1\rangle=|n\rangle\otimes| \da\rangle$.
  The Fock states $|n\rangle$ are  eigenstates of $\ad a$: $\ad a|n\rangle=n|n\rangle$. 
$\Hpm$ is isomorphic to $L^2(\Rr)$, i.e. to the span of Fock states $|n\rangle$ for $n\in\Zpl$. 
To define the finite-dimensional approximants $H_\pm^{(N)}$ to $H_R$ in $\Hpm$, 
one uses projection operators $\Prj_\pm^{(N)}$. $\Prj_\pm^{(N)}$ projects onto $\Hpm^{(N)}=$ span$\{|n,\pm(-1)^n\rangle\}_{0\le n\le N}$. 
The truncated parity chain
$\Hpm^{(N)}$ is then isomorphic to $\Cc^{N+1}$ and
the truncated Hamiltonian in $\Hpm^{(N)}$ reads  $H_\pm^{(N)}=\Prj_\pm^{(N)} H_R \Prj_\pm^{(N)}$.
 $H_\pm^{(N)}$  assumes the tridiagonal form 
\beq
H_\pm^{(N)}=M^{\pm}_0=\left(
\begin{array}{lllllll}
\ \ b^\pm_0&\sqrt{a_1}& \ph{a}0 &\cdots & & \\
\sqrt{a_1}&\ \ b^\pm_1& \sqrt{a_2}& & & & \\
   \ph{a}0& \sqrt{a_2} &  &  & & & \\
    \  \ \vdots& & &\ddots & & & \\
     & & & & &\ b^\pm_{N-1}&\sqrt{a_N}\\
     &  & & & &\sqrt{a_N}&\ b^\pm_N  
\end{array}
\right),
\label{matrix}
\eeq
with
\beq
b^\pm_j=j\om\pm(-1)^j\D, \quad a_j=jg^2, \quad j=0,1, \ldots N.
\eeq
To derive a recurrence relation for $\det M_0^\pm$, define matrices $M_j^\pm$ by deleting the first
$j$ rows and columns from $M^\pm_0$. Then  it follows,
\beq
\det M^\pm_j=b^\pm_j\det M^\pm_{j+1}-a_{j+1}\det M^\pm_{j+2}
\label{3-term-recur}
\eeq
for $j=0,\ldots N$, setting $\det M^\pm_{N+1}=1$, $\det M^\pm_{N+2}=0$.
This  three-term recurrence can be turned into a non-linear two-term
recurrence for $G^\pm_j=\det M^\pm_{j+1}/\det M^\pm_j$,
\beq
G^\pm_j=\frac{1}{b^\pm_j-a_{j+1}G^\pm_{j+1}}.
\label{2-term-recur}
\eeq
To compute the resolvent of $H_\pm^{(N)}$, we  set ${M^\pm_0}'=E-H_\pm^{(N)}$ and find for 
$G^\pm_0(E)=\langle 0,\pm1|(E-H_\pm^{(N)})^{-1}|0,\pm1\rangle$
a representation in terms of  a finite continued fraction
\beq
G_0^{\pm}(E)=1|b^\pm_0(E)-a_1|b^\pm_1(E)-a_2|b_2^\pm(E)-\ldots a_N|b^\pm_N(E),
\label{G-cf}
\eeq
where the $b_j^\pm(E),a_j$ are now defined as
\beq
b^\pm_j(E)=E-j\om\mp(-1)^j\D, \quad a_j=jg^2, \quad j=0,1, \ldots N.
\eeq
We drop the parity index from now on and write $G_0^{(N)}(E)$ for the expression (\ref{G-cf}).
$G_0^{(N)}(E)=\det M_1/\det M_0$ has poles at the eigenvalues of $H^{(N)}$, provided that the pole 
(a zero of $\det (E-H^{(N)})$ at $E_n$) is not lifted by a zero of $\det M_1$, 
which could occur
if $\langle 0,\pm1|\psi_n\rangle=0$
for the eigenstate $|\psi_n\rangle$. This does not happen in the present case because
all $a_j$ are nonzero.
Nevertheless, the numerical evaluation of (\ref{G-cf}) is compromised by the rapid decay
of the overlap $\langle 0,\pm1|\psi_n\rangle$ with growing $n$. It becomes apparent in Fig.~1
of \cite{zie}, where the peaks of the spectral density are very small
already for $n=9,10$.

Method B presumes that the poles of $\lim_{N\ra\infty}G_0^{(N)}$ give the 
spectrum of the untruncated Rabi model 
if $G_0^\infty(E)=\lim_{N\ra\infty}G_0^{(N)}$ can be shown to exist for any real $E$.
To prove this,
Ziegler \cite{zie} invokes the main theorem of Pringsheim \cite{per}, 
which states that the
sequence of finite continued fractions
\beq
F_N=\a_1|\b_1-\a_2|\b_2-\a_3|\b_3-\ldots \a_N|\b_N
\label{pring}
\eeq
converges for $N\ra\infty$ to a value $F$ with $|F|\le 1$ 
if $|\b_j|\ge |\a_j|+1$ for all $j=1,2,\ldots$. If the $\a_j,\b_j$ would be 
dimensionless, this theorem could be used to compute the
tail $\xi_n$ of $G^\infty_0$
\beq
\xi_n(E)=a_n|b_n-a_{n+1}|b_{n+1}-\ldots
\eeq
for sufficiently large $n$ and fixed $E$, because  $|b_j|\sim j\om$ and $|a_j|\sim jg^2$ for
$j\ge n\gg 1$ such that
$|b_j|\ge|a_j|+1$ for all $j\ge n$, provided $g^2<\om$.
The convergence of $\xi_n(E)$ then ensures convergence of the full continued fraction $G_0^\infty(E)$
in the generalized sense, which is sufficient to conclude that the distribution of the poles of 
$G^\infty_0(E)$ is well-defined.
 
To obtain dimensionless quantities, $g$ and $\D$ in (\ref{ham1}) are scaled with $\om$: $\tilde{g}=g/\om$, $\tilde{\D}=\D/\om$, 
$\tilde{\om}=1$.
The condition $g^2<\om$ turns into ${\tilde{g}}^2<1$, or $|g|<\om$. This condition
seems to invalidate the proof for the deep strong coupling regime $|g|\gtrsim \om$, which is of current interest \cite{sol}. 
But Ziegler \cite{zie} does not rescale the Hamiltonian and obtains instead of $|g|<\om$ the meaningless
expression $g^2<\om$, which compares quantities of different dimension. However, it is easy to 
correct this mistake by using the following
generalization of Pringsheim's theorem for dimensionful quantities which lifts at the same time the restriction $|g|<\om$ \cite{note3}.\par
\vspace{3mm}
\noindent
Theorem:\ Consider the continued fractions (\ref{pring}) and assume for the dimension $[F_N]$ of $F_N$ the
relation $[F_N]=[\b_j]$ and $[\a_j]=[F_N^2]$. If $\prod_{j=1}^n|\a_j|$ is unbounded in $n$ and
the inequalities $|\b_j|\ge |\a_j|/c +c$ hold for all $j\ge 1$ and some constant $c$
with dimension $[c]=[\b_j]$, then $\lim_{N\ra\infty}F_N=F$ exists and $|F|\le c$.
\par
\vspace{3mm}
The theorem is applicable to the tail $\xi_n(E)$ of $G^\infty_0(E)$. Moreover, the possibility to choose $c$
freely leads to the following lower bound for $n$,
\beq
n \ge \frac{|E|+|\D|}{\om}+\frac{2g^2}{\om^2}\left(1+\sqrt{1+\frac{(|E|+|\D|)\om}{g^2}}\right). 
\label{bound}
\eeq  
In this way, convergence of $G^\infty_0(E)$ can be proven for arbitrary model parameters. (An alternative
would be to apply the  corollary used by Schweber \cite{note6}). 

I shall now demonstrate that this proof of convergence is not sufficient to show that $\lim_{N\ra\infty}G_0^{(N)}(E)$ is related
to the spectrum of the full Rabi model for fixed parity. 
The reason is the intrinsic ambiguity in the definition of the finite-dimensional approximants to $H_R$. 
If $H_R$ were compact, it could be defined as the limit $\lim_{N\ra\infty}H^{(N)}$ 
in the norm topology \cite{note4}
 and $\s(H^{(N)})$ would converge towards $\s(H_R)$  in the sense defined in (\ref{cond-lim}) because the sets $\{E^{(N)}_n\}$ and $\{E_n\}$ would satisfy the even stronger condition
\beq
\lim_{N\ra\infty}\sup_{n\in \Zpl_0}|E_n^{(N)}-E_n|=0.
\eeq  
But $H_R$ is unbounded in ${\cal H}$ and therefore not the limit of any sequence of finite-dimensional operators.
Even if the sequence of truncated state spaces $\{{\cal H}^{(N)}\}_{n\in\Zpl}$ could be determined in a unique way (which is not the case), the approximating operators $H^{(N)}$ need not to be the $H^{(N)}=\Prj^{(N)}H_R\Prj^{(N)}$, used in method B. 
The projection $\Prj^{(N)}$ affects only the states $|N\rangle$ and $|N+1\rangle$ because it is equivalent to setting
$\langle N|H_R|N+1\rangle=|g|\sqrt{N+1}$ to zero. But as these matrix elements grow as $\sqrt{N}$ for $N\ra\infty$, it is not clear whether the  spectra $\s(H^{(N)})$ converge to $\s(H_R)$, even if they do converge to a limit. One could think of other modifications of the matrix elements of $H_R$ between states with high energy and claim that these are also 
admissible truncation prescriptions.
A similar ambiguity is well-known from quantum field theory: The regularization scheme needed to make the perturbative expressions finite by truncating the ultra-violet modes is not uniquely definable. One has to prove that the resulting (renormalized) expressions are {\it independent} from the choice of cut-off. 
In our case, the task is to show that the spectra of a 
sequence of finite-dimensional
Hamiltonians $\{H^{(N)}\}$ converge to  $\s(H_R)$, and to this end 
one would have to prove that this limit does not depend on the modification of $H_R$ in the high-energy region which is used to define the approximants $H^{(N)}$.
The formal condition reads as follows:

If the elements of two sequences of operators $\{H^{(N)}\}$ and $\{\sH^{(N)}\}$ are defined on the spaces ${\cal H}^{(N)}$ for
$N=0,1,\ldots$ with ${\cal H}^{(0)}\subset{\cal H}^{(1)}\subset \ldots L^2(\Rr)$ and $H^{(N)}=\sH^{(N)}=H_R$
if projected onto ${\cal H}^{(N-1)}$, the limit of the spectra of both sequences  as defined in (\ref{cond-lim}) must coincide: 
$\lim_{N\ra\infty}\s(H^{(N)})=\lim_{N\ra\infty}\s(\sH^{(N)})$.
 
This is obviously necessary (albeit not sufficient) to justify the claim that $\s(H^{(N)})$ converges to the unique $\s(H_R)$, if $H_R$ is
not compact and the $H^{(N)}$ do not converge to it in the operator sense, 
but it is violated in the present case:  
One may define a cut-off prescription for $\sH^{(N)}$
which affects only the states $|N-1\rangle$, $|N\rangle$ and $|N+1\rangle$ but produces a (fictitious) pole in $G_0^{(N)}(E)$
at an arbitrary value $E_0$. Consider the expression (\ref{G-cf}) for  $G_0$, which reads
\beq
G_0(E) = 1|b_0(E)-a_1|b_1(E)-a_2|b_2(E) 
\ldots\-  a_{N-1}|b_{N-1}-a_NG_N(E)
 \eeq
as follows from (\ref{2-term-recur}). The standard cut-off using the projection $\Prj^{(N)}$ sets $G_N(E)=1/b_N(E)$.
But $G_N(E)$ can also be determined from below by a recurrence relation equivalent to (\ref{2-term-recur}),
\beq
G_{j+1}(E)=\frac{b_j(E)}{a_{j+1}}-\frac{1}{a_{j+1}G_j(E)}.
\label{inv-recur}
\eeq 
If one sets $G_0(E_0)=\infty$, corresponding to a pole of $G_0(E)$ at $E_0$,
the recurrence (\ref{inv-recur}) with the initial condition $G_1(E_0)=b_0(E_0)/a_1$
yields a $G_N(E_0)$ which will tend to the value $-\om/g^2$ for  $N\ra\infty$.
The matrix element $\langle N|\sH^{(N)}|N\rangle$ of the truncated Hamiltonian  $\sH^{(N)}$ is defined by
\beq
\langle N|\sH^{(N)}|N\rangle=\sH^{(N)}_{N,N}=E_0-\frac{1}{G_N(E_0)}.
\eeq
For all $(i,j)\neq(N,N)$,  $H_{ij}^{(N)}=\sH_{ij}^{(N)}$. But the ensuing
$\tilde{G}_0^{(N)}(E)$ will have a pole at $E_0$, although only a single matrix element 
was changed and therefore $\Prj^{(N-1)}\sH^{(N)}\Prj^{(N-1)}=\Prj^{(N-1)}H_R\Prj^{(N-1)}$. The same result follows if one sets
$\sH_{N,N}^{(N)}=E_0-N/G_N(E_0)$ and modifies at the same time  $\sH^{(N)}_{N-1,N}
=gN$, which
does not yield a low energy state {\it a priori} if $G_N+\om/g^2$ tends slower to zero  than $1/N$.
It is therefore possible to define a sequence of 
Hamiltonians $\{\sH^{(N)}\}$  with  eigenvalues which are not close to elements of the spectrum of the untruncated Rabi model, but differ from the $H^{(N)}$  in at most three matrix elements and coincide with $H_R$ on ${\cal H}^{(N-1)}$.
Neither  $\{\sH^{(N)}\}$ nor  $\{H^{(N)}\}$ converge to $H_R$ in norm topology because $H_R$ is unbounded. This entails that no independent criterion exists to select the ``correct'' sequence  $\{H^{(N)}\}$ apart from
convergence of the spectrum in the sense of (\ref{cond-lim}) to some set $\{e_n\}_{n\in\Zpl_0}$. However, the spectra of the  sequence $\{\sH^{(N)}\}$
converge as well, if
$\sH^{(N)}_{N,N},\sH^{(N)}_{N,N-1}$ are appropriate functions of $N$. The problem cannot be solved by consideration of resolvents instead of the Hamiltonians. Whereas $H_R$ has compact resolvent, the resolvents of the finite-range operators $H^{(N)}$ and $\sH^{(N)}$ are bounded but not compact and do not converge to
$(z-H_R)^{-1}$ in norm topology. Even if one could prove that 
only $\{H^{(N)}\}$ and not  $\{\sH^{(N)}\}$ converges to 
$H_R$ in strong resolvent sense, it would be insufficient 
to determine $\s(H_R)$ unambiguously \cite{note5}.    

The standard cut-off prescription leads indeed to the correct spectrum of the full model as the comparison
with the analytical solution \cite{db} shows. But  this prescription is only valid because
the finite approximants $H_N$ obtained by method A are the projections
$\Prj_NH_R\Prj_N$ and thus correspond to the standard cut-off of the model
in the finite-dimensional state spaces ${\cal H}_N$ as shown above. These spaces are not generated by the first $N+1$ Fock states as the ${\cal H}^{(N)}$ but are sufficiently close to them for large $N$ 
because the coherent factor $e^{-gz/\om}$ appearing in the $\psi(z;N)$ has a convergent
expansion in Fock states. The isomorphism $\hat{I}$ between  $\B$ and $L^2(\Rr)$ maps the function $z^n/\sqrt{n!}$ onto 
$|n\rangle$; it follows that for any vector $|\psi_N\rangle\in \hat{I}({\cal H}_N)\subset L^2(\Rr)$ and its projection $|\psi_N'\rangle$ onto ${\cal H}^{(N)}$, we have
\beq
\lim_{N\ra\infty}\langle\psi_N-\psi_N'|\psi_N-\psi_N'\rangle =0.
\eeq
Therefore $\Prj_NH_R\Prj_N-\Prj^{(N)}H_R\Prj^{(N)}$ tends to zero in the norm topology for $N\ra\infty$.
 We may conclude that method B yields the  spectrum of the Rabi model
in the limit $N\ra\infty$ precisely because the finite-dimensional 
approximants $H_N$ of method A are the correct ones. Method B cannot be justified independently
from method A \cite{note7}.    

Apart from the implicit dependence on method A, 
method B shares with it the same problems: The expressions derived from continued fractions do not allow for {\it qualitative}
analysis because 
the singularities of
neither $F_\infty(E)$ nor $1/G_0^\pm(E)$ are known, which would be necessary to infer the distribution of the eigenvalues of $H_R$. Moreover, the position and nature of spectral degeneracies cannot be obtained from those functions, as
exemplified in \cite{zie}, where the following is said
on level crossings:
``The individual matrix elements $g,h$  avoid level crossing due to parity conservation,
since eigenstates of consecutive eigenvalues have different parity.'' 
However, $g$ and $h$ correspond to the subspaces with even and odd parity, respectively; the parity does {\it not} change 
within each subspace. The idea that alternating parity forbids level crossing 
between consecutive eigenvalues is erroneous, because  the converse is true \cite{db}.
If applied to the full Hilbert space, Zieglers statement is wrong as well: 
Energetically neighboring states do not have always different
parity. Finally, the fact that
the degeneracies (of states with different parity) occur only if $E+g^2/\om$
is an integer multiple of $\om$ 
can be deduced neither with method A nor method B.

\section{Conclusions}

We conclude that both methods A and B yield correct approximations for 
the spectrum of the quantum Rabi model and are equivalent to numerical
exact diagonalization in finite-dimensional Hilbert spaces of sufficient large dimension. Diagonalization on a truncated state space
has been used for many years
as a tool to obtain numerically exact results which were subsequently compared to a variety of analytical approximations (see e.g. \cite{irish}) without questioning its correctness, although the Hamiltonian
$H_R$ is not the limit of a sequence of finite-dimensional operators. The soundness of the approach was taken for granted because of numerical convergence of the results, which is equivalent with  convergence of the corresponding continued fraction.
The pointwise convergence of the latter has been established by Schweber
\cite{schweb}
for method A and by Durst~{\it et al.} \cite{durst} for method B. In the case of method A, this convergence
is equivalent to the analyticity condition for the {\it untruncated} model and therefore sufficient to prove numerical identity with the exact spectrum
for large enough $N$.
In contrast to method A, the convergence of the continued fraction says
nothing about the spectrum of the full model within the framework of method B. The sequence
of approximating Hamiltonians $H^{(N)}$ is not uniquely determined and spectral
convergence alone is not sufficient to prove a connection with $H_R$.
But because the cut-off scheme employed by method B is equivalent (for large $N$) to the
sequence of finite-dimensional approximants $H_N$ from method A, it yields
the (numerically) exact spectrum in the limit $N\ra\infty$ as well. It owes
this to the validity of method A: Contrary to the claim made in \cite{zie},
method B cannot be considered as an independent way to obtain the
correct spectrum of $H_R$.

In this way the use of continued fractions and (equivalently)
numerical exact diagonalization can be rigorously justified
in the case of the Rabi model. The key to the proof is Bargmanns
analyticity condition for functions as elements of an infinite-dimensional 
Hilbert space. Whether this equivalence with the analytical solution
can be extended to more complicated systems or breaks down in some cases
\cite{trav} should be the subject of future study.

 \acknowledgments
 This work was supported by Deutsche Forschungsgemeinschaft through TRR~80.


\begin{thebibliography}{99}
\bibitem{note1} One assumes here that the minimal solution has $K_0\neq 0$, 
which does not follow from (\ref{recur}). It is therefore not clear {\it a priori} that 
$K^{min}_1$ is related for all $E$ to 
the continued fraction (\ref{cont-frac2}), which means that the convergence of
(\ref{cf-limit}) has to be proven independently.
\bibitem{note7}
While an independent justification cannot be excluded in principle, it seems likely that it will involve sophisticated mathematical arguments, which, to the best of my knowledge, have not been presented yet.
\end{thebibliography}
\end{document}